\def \cm{~\rm{cm}}
\def \s{~\rm{s}}

\def \erg{~\rm{erg}}

\documentclass[12pt,preprint]{aastex}

\shorttitle{Symbiotic Nebulae and the PNLF}
\shortauthors{Frankowski \& Soker}

\begin{document}

\title{COMPARING SYMBIOTIC NEBULAE AND PLANETARY NEBULAE LUMINOSITY FUNCTIONS}
\author{Adam Frankowski \& Noam Soker \altaffilmark{}}
\affil{Department of Physics, Technion$-$Israel Institute of
Technology, Haifa 32000 Israel; adamf@physics.technion.ac.il;
soker@physics.technion.ac.il}

\begin{abstract}
We compare the observed symbiotic nebulae (SyN) luminosity function (SyNLF)
in the [O~III] $\lambda 5007~$\AA~ line to the planetary nebulae (PN)
luminosity function (PNLF) and find that the intrinsic SyNLF (ISyNLF)
of galactic SyNs has
-- within its uncertainty of 0.5--0.8mag --
very similar cutoff luminosity and general shape to those of the PNLF.
The [O~III]/(H$\alpha$+[N~II]) line ratios of SyNs and PNs are shown to be also
related. Possible implications of these results for the universality of the
PNLF are briefly outlined.

\end{abstract}

\section{INTRODUCTION}
\label{sec:intro}

The luminosity of emission lines in a planetary nebula (PN) depends on
the mass and evolutionary status of the central star (CSPN), which determine
the temperature and luminosity of the CSPN, and on the nebular properties.
These in turn depend on the initial mass of the progenitor, on its metallicity,
and most likely on interactions with a binary companion.
Therefore, the PN luminosity function (PNLF) depends on the age of the parent
population (Dopita et al. 1992; M\'{e}ndez et al. 1993; Marigo et al. 2004),
and, to lesser degree, on the metallicity (Ciardullo \& Jacoby 1992; Marigo et al. 2004),
hence it varies between different types of galaxies (e.g., Ciardullo et al. 2004).
However, the [O~III]~$\lambda5007~$\AA~ (hereafter [O~III]) most luminous end
of the PNLF, and in particular the cutoff (maximum) [O~III] luminosity,
seem to be similar in all large PN populations, with very small dependance
on galaxy type (Ciardullo et al. 2005).
This allows a very successful use of the PNLF as a standard candle, from the
galactic bulge (Pottasch 1990), through the LMC and SMC (Jacoby et al. 1990),
M31 (Ciardullo  et al. 1989) and to galaxies at larger distances,
spirals (e.g.  Feldmeier et al. 1997), and ellipticals (Jacoby et al. 1996).

Some basic properties of the PNLF are well understood (e.g., Dopita et al. 1992;
M\'{e}ndez et al. 1993; Marigo et al. 2004; review by Ciardullo 2006a).
Recently, further progress on the underlying physics has been brought
through hydrodynamical PN modeling (Sch\"onberner et al.~2007).
However, a major puzzle is the cutoff (maximum) luminosity in old stellar populations,
such as in elliptical galaxies, which is the same as in young populations
(Ciardullo et al. 2005; Ciardullo 2006a).
This cutoff luminosity of the PNLF is $L$[O~III]~$\simeq 600 L_\odot$,
and it leads to a chain of constraints (Ciardullo et al. 2005; Ciardullo 2006a):
Such [O~III] luminosity requires the ionizing central star to have a bolometric
luminosity of $L_\ast \ga 6000 L_\odot$, which in turn requires the
central star to have a mass of $M_\ast > 0.6 M_\odot$, and the progenitor
to have a main sequence mass of $M> 2 M_\odot$.
Such a progenitor mass is not expected in old stellar populations.
Single star evolution alone cannot account for this finding (Marigo et al. 2004).

Three explanations to the universality of the the [O~III] cutoff have been proposed.
($i$) Ciardullo et al. (2005) proposes that the most [O~III] luminous PNs in old
stellar populations are descendant of blue-straggler type stars; namely, two lower
mass stars, $\sim 1 M_\odot$, merged on the main sequence to form a star of
mass $\sim 2 M_\odot$.
($ii$) Soker (2006) proposed that most, or even all, of the [O~III] luminous PNs in old
stellar populations are actually evolved symbiotic nebulae (SyNs).
Ciardullo (2006b) raised some problems with this explanation.
Our reply to these will be discussed in a forthcoming paper.
In this paper we limit ourselves to presenting an interesting similarity
between the PNLF and the symbiotic nebulae luminosity function (SyNLF).
($iii$) It is possible that low mass stars in old stellar population can form
massive ($\sim 0.63 M_\odot$) PN central stars (CSPN ), as mass loss rate is lower for
low metallicity stars (see discussion in M\'{e}ndez et al. 2008).

\section{COMPARING THE LUMINOSITY FUNCTIONS}
\label{sec:basic}

\begin{figure}
\hspace{2cm}
\resizebox{0.7\hsize}{!}{\includegraphics{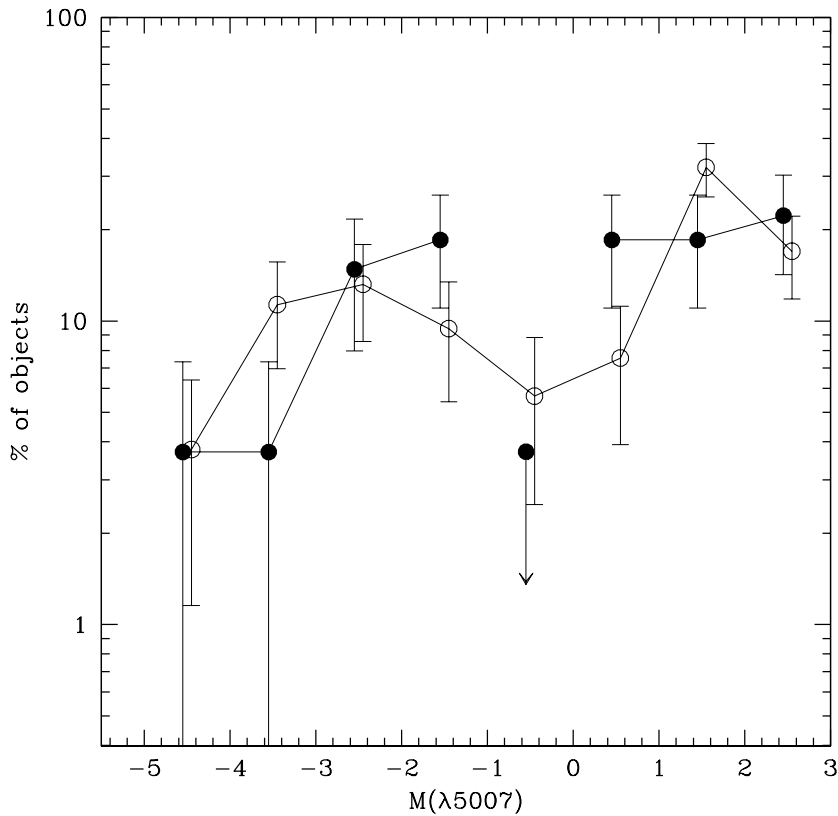}}
\caption{The intrinsic luminosity function of symbiotic nebulae (ISyNLF)
as calculated from the sample given by M97 (filled
circles) compared to the planetary nebula luminosity function (PNLF) given
by (Jacoby \& De Marco 2002; open circles).
The points are slightly displaced from the centers of the bins to increase
readability.
Error bars from the binomial distribution are plotted.
Both distributions are normalized to correspond to the same sample size
for $M(\lambda 5007) > 3$mag.
There are 53 PNs and 27 SyNs in the samples used for the plots;
the leftmost point has one object in the SyN sample and 2 objects in the PN sample.
The empty [-1,0) bin for the symbiotic sample is denoted by an
upper limit set at 1/(sample size).}
\label{fig:lf}
\end{figure}

To compare SyNs with PNs we analyze the data
from Miko{\l}ajewska et al.~(1997; hereafter M97), who provide emission
line fluxes for a sample of 67 southern SyNs, resulting from a survey
of objects classified as PNs in various catalogs.
We only use objects from this dataset, because it is
the only one available in the literature that combines the advantages
of a sizeable sample and uniform treatment with an extensive coverage
of data relevant for such a comparison.
M97 give also $E(B\!-\!V)$ towards the nebulae, based
on their own data as well as data gathered from the literature.
H$\alpha$ fluxes are available for all the objects in the catalog,
and 41 SyNs also have measurements for the [O~III] line and a
known distance.
This final sample contains one D'-type object, 10 D-type, and 30
S-type, giving a fair representation of the frequencies among
galactic symbiotics (e.g., Belczy{\'n}ski et al. 2000, Miko{\l}ajewska 2003).
For these objects we reconstruct the intrinsic nebular fluxes
in the H$\alpha$ and $5007$\AA~ lines using the extinction curve of
Seaton (1979), which was also assumed by M97. From
the dereddened [O~III] fluxes and distances we build the {\em intrinsic}
luminosity function (ISyNLF) of this SyN sample in $M(\lambda 5007)$,
the absolute magnitude of the [O~III] line, employing the definition
of the apparent magnitude in this line from Jacoby (1989):
\begin{equation}
m(\lambda 5007) = -2.5 \log F(\lambda 5007) -13.74
\end{equation}
where the flux $F$ is in $\erg \cm^{-2} \s^{-1}$.
Taking into account the uncertainties in the distances
and in the $E(B\!-\!V)$ towards the nebulae, we estimate the statistical
errors of our computed $M(5007)$ values to be 0.5--0.8mag.
In Table~\ref{tab:top6} we list the SyNs with $M(\lambda 5007)<-2$mag
(that occupy the three brightest bins in Fig~\ref{fig:lf}).
Not surprisingly, all of them are D-type objects (i.e., containing a Mira).
\begin{table}
\caption{Parameters of the SyNs with $M(\lambda 5007)<-2$mag, i.e., up to
2.5mag from the standard PNLF cutoff, $M^*\!\approx\!-4.5$mag
(contained in the three top bins of the ISyNLF in Fig~\ref{fig:lf}).
Observed fluxes are in units of $10^{-13}$ erg s$^{-1}$ cm$^{-2}$.}
\label{tab:top6}
\medskip
\begin{tabular}{lr@{.}lr@{.}lcr@{.}lcc}
\hline
\hline
\multicolumn{1}{c}{Name} & \multicolumn{2}{c}{$F({\rm H}\alpha)^a$} \
& \multicolumn{2}{c}{$F(\lambda 5007)^a$} \
& $E(B\!-\!V)^a$ & \multicolumn{2}{c}{$d^a$} \
& $M(\lambda 5007)$ \
& $I(\lambda 5007)$ \\
\cline{10-10}
 & \multicolumn{2}{c}{} & \multicolumn{2}{c}{} \
&  & \multicolumn{2}{c}{[kpc]} \
& mag & $I({\rm H}\alpha)$ \\
\hline
V835 Cen    & 411&40 &  \phantom{m}66&30 & 2.0 &  1&9 & -4.29 & 1.14 \\
H2-38       & 179&70 &  33&70 & 1.1 &  7&2 & -3.25 & 0.55 \\
He2-171     &  47&60 &  12&60 & 1.4 &  5&0 & -2.46 & 1.04 \\
V704 Cen    &   7&20 &   0&88 & 1.5 & 16&0 & -2.45 & 0.53 \\
He2-38      & 518&80 &  44&90 & 1.5 &  2&2 & -2.41 & 0.37 \\
V852 Cen$^b$& 215&43 &  57&57 & 1.0 &  4&3 & -2.36 & 0.71 \\
\hline
\end{tabular}

\medskip
$^a$ From Miko{\l}ajewska et al.~(1997)\\
$^b$ Fluxes shown for this object are averages from 3 measurements
\end{table}

The SyN sample is quite small and yet it spans $\sim 10$mag in $M(\lambda 5007)$,
so a comparison with a reasonably deep PNLF is preferred.
Figure~\ref{fig:lf} compares the ISyNLF in the [O~III]
line to the deep PNLF of the SMC (Jacoby \& De Marco 2002),
assuming 19.3mag distance modulus to SMC and galactic extinction
of 0.2mag (M\'{e}ndez et al. 2008;
we note that the extinction value is deduced by us from their fig.~9,
as they do not give it explicitly).
The binning in the graphs is coarse because of the low number of objects
(53 PNe and 27 SyN brighter than 3mag), and the error bars are large.
Nevertheless, the two luminosity functions
are consistent at all points within their binomial errors and
are remarkably similar in terms of the maximum
brightness cutoff and general shape, including a dip around $-1$--0mag.

\begin{figure}
\hspace{2cm}
\resizebox{0.7\hsize}{!}{\includegraphics{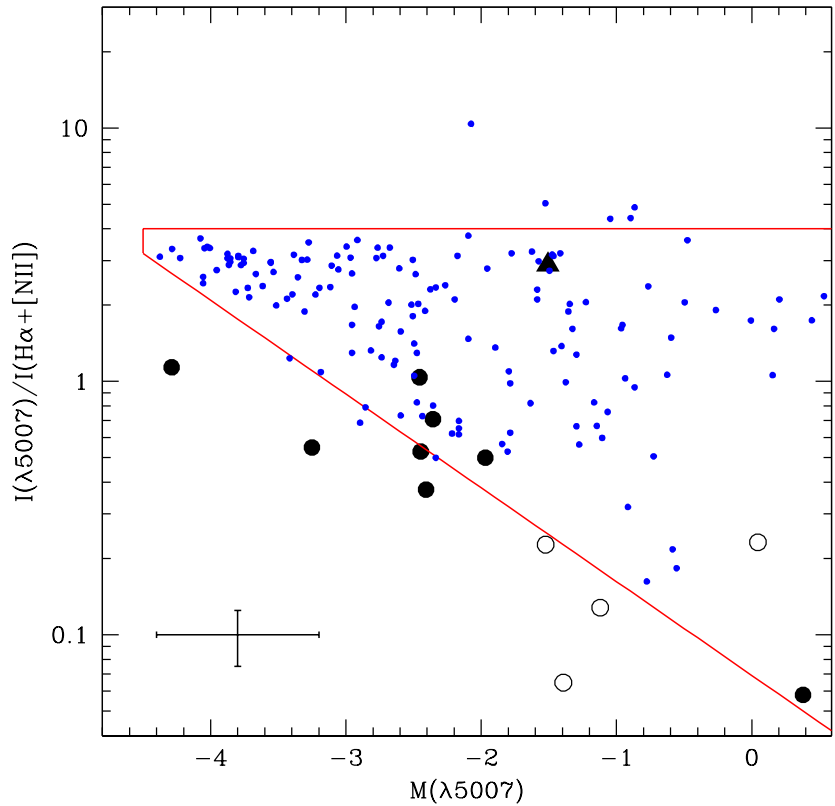}}
\caption{The [O~III] $\lambda$5007 to H$\alpha$ line ratios for the
SyN sample used in this paper (large symbols) compared with the
[O~III] $\lambda$5007 to H$\alpha$+[N~II] line ratios for PNs in M33
as given by Ciardullo (2004; dots).
Different symbols mark IR classification from M97 (filled circles:
D-type objects, open circles: S-type, filled triangle: D'-type).
The cross in the lower left corner shows typical error bars in the SyN
sample.
The contour line delineates the PNs populated region, observationally
found constant across various galaxies, as defined in Herrmann et al. (2008).
Notice that for a clear comparison with PNs the plot is limited to the
$M(\lambda 5007)$-bright region and only the brightest 13 SyNs
are visible here; the rest lies to the right of the displayed region.
}
\label{fig:ratio}
\end{figure}

The intensity ratio between the [O~III] and H$\alpha$+[N~II] emission lines is
an important parameter in identifying extragalactic PNe
(e.g., Ciardullo et al. 2002).
The [N~II] lines are located $\sim 20$\AA~ to the sides of H$\alpha$ and so
are not separated in narrow-band photometry used to find extragalactic PNs.
They would be easily resolved in the spectroscopic observations
of M97, but the [N~II] lines in SyNs are with few
exceptions either not observed or very weak compared to H$\alpha$
(e.g., Van Winckel et al. 1993; Ivison et al. 1994), so for the SyNs
the distinction between H$\alpha$ and H$\alpha$+[N~II] intensities is
not essential and we use the H$\alpha$ fluxes provided by M97.
Figure~\ref{fig:ratio} plots the [O~III]/H$\alpha$ line ratio for the
SyNs and the [O~III]/(H$\alpha$+[N~II]) ratio for M33 PNs
(Ciardullo et al. 2004) vs. the absolute [O~III] magnitude, $M(\lambda 5007)$.
PNs in various galaxies consistently populate the area to the right
of the contour line, defined in Herrmann et al. (2008).
The difference in position between PNs and SyNs is clear, as
is the similarity of the shape of the occupied regions.
The SyN region is basically displaced with respect to the PN region
by $\sim 0.5$dex along the line-ratio axis.

\section{DISCUSSION AND SUMMARY}
\label{sec:discussion}

Even though we use intrinsic properties for the SyNs
and measured properties for the extragalactic PNs, the comparison --
though crude -- is readily justified, as follows.
The SyNs lie in the galactic plane and suffer large reddening:
the characteristic $E(B\!-\!V)$ of SyNs in the sample of M97 is 1--2mag.
Not having separate information about interstellar and circumstellar
extinction, we use the total reddening values and plot the properties
the SyNs would have if completely unobscured.
For extragalactic PNs the plots presented in the
literature take into account the Milky Way reddening and
the estimated reddening within the host galaxy (not always),
but not their intrinsic circumstellar extinction, which is usually unknown.
However, the typical circumstellar contribution to $E(B\!-\!V)$ for PNs is
0.0--0.2mag,
and only in rare cases up to $\sim 0.5$mag (Ciardullo \& Jacoby 1999).
The extragalactic PN samples would need relatively little additional
de-reddening to be brought to the intrinsic values.
Therefore the comparison is more sound than it might seem at a first glance.

Ciardullo (2006b) shows a difference in the diagrams of line ratio vs.
[O~III] magnitude of PNs in the LMC and in M31 that results from inclusion
of the intrinsic reddening.
Accounting for the intrinsic obscuration does shift the observed
bright tip of the distribution by about 0.8mag, to $\sim 0.5$mag above
the standard PNLF cutoff value, $M^* \approx -4.5$mag.
However, given our symbiotic sample size, this should not be taken as
evidence of a strong difference in the intrinsic $M(\lambda 5007)$ between
PNs and SyN, especially since Corradi \& Magrini (2006) show that in the Local
Group galaxies with PN samples of comparable size the brightest observed PNs
can easily fall $\sim 1$mag below the cutoff.

\begin{figure}
\hspace{2cm}
\resizebox{0.7\hsize}{!}{\includegraphics{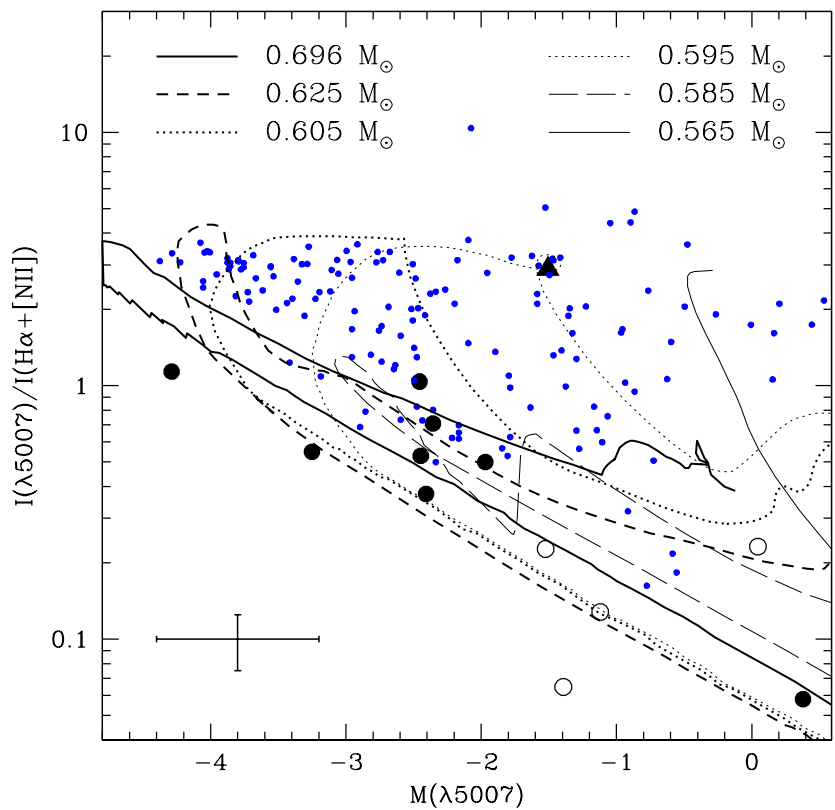}}
\caption{Same as Fig.~\ref{fig:ratio} but with PN evolution tracks
from M{\'e}ndez et al. (2008) overplotted. The models show evolution of
PNs around post-AGB cores of various masses.
Notice how SyNs cover the lower (ascending) portions of the tracks,
where PNs are lacking. Notice also the location of the
D'-type system HD~330036 among the upper branches of the tracks.
}
\label{fig:ratio2}
\end{figure}

Interestingly, M\'{e}ndez et al.~(2008) plot the theoretical evolutionary
tracks of PN models from Sch\"onberner et al.~(2007)
in the [O~III]/(H$\alpha$+[N~II]) vs. $M(\lambda 5007)$
plane and find that the rising portions of these tracks cluster
{\em below} the observed locus of PNs, exactly in the area which in the
present paper is shown to be occupied by SyNs (see Fig.\ref{fig:ratio2}).
M\'{e}ndez~et~al.~(2008) explain the lack of PNs in this area by
the relatively fast evolution of PNs through this region, but it should
be stressed that their theoretical tracks display {\em intrinsic} nebular
properties, as no internal reddening was applied to the tracks.
Strong intrinsic reddening (as is arguably the case for both SyNs and
proto-PNs)
removes the objects from this region -- they can be compared to the
theoretical tracks only when dereddened, as we did here for the SyNs.
Notice that in the scenario of Soker (2006) the contamination
of the {\em observed} PNLF by SyNs would most likely occur in
the post-symbiotic stage, after the donor has left the AGB, its massive
wind ceased, and the intrinsic extinction dropped -- but the hot accretor
is still the main ionizing source.
%

The fact that the top of the ISyNLF is dominated by D-type systems
is understandable: Mira winds can provide both abundant fuel for nuclear
reactions on the accreting white dwarf (WD) and a massive nebula for
it to ionize.
Although four objects in Figures~\ref{fig:ratio}--\ref{fig:ratio2}
are classified as S-type in M97, every one of them have been
considered D-type by some authors (see M97, Belczy{\'n}ski et al. 2000).
The undisputed S-type objects are all dimmer than
$M(\lambda 5007) = 0.5$mag.
Therefore, the gap in the ISyNLF around $-1$--$0$mag marks the division
between the D-type and the S-type systems. Their separation reflects
the fact that in SyNs the evolution of the hot component and of
the nebula is forced by the mass loss from the giant.
It is the D-type objects that are relevant when comparing
the upper ISyNLF to the tip of the PNLF.
And it is these systems, according to the scenario of Soker (2006),
that will consequently contaminate the observed PNLF in their post-symbiotic
stage.
The only D'-type object in the analyzed sample, HD~330036, falls well
into the PN region on the line ratio vs. [O~III] magnitude diagram
(Fig.~\ref{fig:ratio2}).
This is consistent with the mounting evidence confirming the
long-lasting suspicion that the D'-type represents in fact young PNs
with a late-type companion (Jorissen et al. 2005).

The similarity between the SyNLF and the PNLF might have different
explanations and implications. In this paper we limit ourselves to listing
a few of the possibilities:
(1) The similarity is a coincidence.
(2) The similarity can be explained by the similarity of the ionizing source,
a nuclear burning on a compact WD (or a WD in the making), and the similarity
of the ionized material, an expanding dense wind of a red giant (RGB or AGB).
Binarity of CSPNs plays no role.
(3) The similarity results from binarity as well as from the similar ionizing
source and a similar origin of nebulae material.
In this explanation, the bright PNs owe their brightness, among other things,
to a strong binary interaction.
We postpone detailed study and discussion of these points to a forthcoming paper.

As a final note, we would like to stress the need
for a dedicated search for SyNs in the Magellanic Clouds (MC).
Because of their proximity and known distance, MC would be ideal
for a study of the PN and SyN population properties in the same
environment, while at the same time sidestepping the issue
of galactic extinction.
Unfortunately, the sample of presently known MC symbiotics
is very small (e.g., Belczy{\'n}ski et al. 2000) and, to the best
of our knowledge, their line fluxes have not been published.
A systematic observational effort is required to enable a direct
comparison of the MC PN and SyN luminosity functions.

\acknowledgments
The authors are very grateful to Joanna Miko{\l}ajewska for valuable
discussions, and to Romano L. M. Corradi, George H. Jacoby, Joel H. Kastner,
Roberto H. M\'{e}ndez, Detlef Sch\"onberner, Romuald Tylenda,
and the anonymous referee for their helpful comments.
This research was supported in part by the Asher
Fund for Space Research at the Technion, and the Israel Science
Foundation.
The authors acknowledge the support of the B.~and G.~Greenberg Research
Fund (Ottawa).

\end{document}